\documentclass[preprint,aps]{revtex4}
\usepackage{amssymb,epsf}

\begin{document}

\title{Thermodynamics of Asymptotically Flat Charged Black Holes in Third Order
Lovelock Gravity}
\author{M. H. Dehghani$^{1,2}$\footnote{email address: mhd@shirazu.ac.ir} and M. Shamirzaie$^{1}$}
\address{$^1$Physics Department and Biruni Observatory, College of Sciences,
Shiraz University, Shiraz 71454, Iran\\ $^2$Research Institute for
Astrophysics and Astronomy of Maragha (RIAAM), Maragha, Iran}

\begin{abstract}
We present a new class of asymptotically flat charge static
solutions in third order Lovelock gravity. These solutions present
black hole solutions with two inner and outer event horizons,
extreme black holes or naked singularities provided the parameters
of the solutions are chosen suitable. We find that the uncharged
asymptotically flat solutions can present black hole with two
inner and outer horizons. This kind of solution does not exist in
Einstein or Gauss-Bonnet gravity, and it is a special effect in
third order Lovelock gravity. We compute temperature, entropy,
charge, electric potential and mass of the black hole solutions,
and find that these quantities satisfy the first law of
thermodynamics. We also perform a stability analysis by computing
the determinant of Hessian matrix of the mass with respect to its
thermodynamic variables in both the canonical and the
grand-canonical ensembles, and show that there exists only an
intermediate stable phase.
\end{abstract}

\maketitle

\section{Introduction}

Both string theory as well as brane world cosmology assume that spacetime
possesses more than four dimensions. In string theory, extra dimensions were
promoted from an interesting curiosity to a theoretical necessity since
superstring theory requires a ten-dimensional spacetime to be consistent
from the quantum point of view, while in the brane world cosmology (which is
also consistent with string theory) the matter and gauge interaction are
localized on a 3-brane, embedded into a higher dimensional spacetime and
gravity propagates in the whole of spacetime. This underscores the need to
consider gravity in higher dimensions. In this context one may use another
consistent theory of gravity in any dimension with a more general action.
This action may be written, for example, through the use of string theory.
The effect of string theory on classical gravitational physics is usually
investigated by means of a low energy effective action which describes
gravity at the classical level \cite{Wit1}. This effective action consists
of the Einstein-Hilbert action plus curvature-squared terms and higher
powers as well, and in general give rise to fourth order field equations and
bring in ghosts. However, if the effective action contains the higher powers
of curvature in particular combinations, then only second order field
equations are produced and consequently no ghosts arise \cite{Zw}. The
effective action obtained by this argument is precisely of the form proposed
by Lovelock \cite{Lov}:

\begin{equation}
I_{G}=\int d^{d}x\sqrt{-g}\sum_{k=0}^{[d/2]}\alpha _{k}\mathcal{L}_{k}
\label{Lov1}
\end{equation}
where $[z]$ denotes integer part of $z$, $\alpha _{k}$ is an arbitrary
constant and $\mathcal{L}_{k}$ is the Euler density of a $2k$-dimensional
manifold,
\begin{equation}
\mathcal{L}_{k}=\frac{1}{2^{k}}\delta _{\rho _{1}\sigma _{1}\cdots \rho
_{k}\sigma _{k}}^{\mu _{1}\nu _{1}\cdots \mu _{k}\nu _{k}}R_{\mu _{1}\nu
_{1}}^{\phantom{\mu_1\nu_1}{\rho_1\sigma_1}}\cdots R_{\mu _{k}\nu _{k}}^{%
\phantom{\mu_k \nu_k}{\rho_k \sigma_k}}  \label{Lov2}
\end{equation}
In Eq. (\ref{Lov2}) $\delta _{\rho _{1}\sigma _{1}\cdots \rho _{k}\sigma
_{k}}^{\mu _{1}\nu _{1}\cdots \mu _{k}\nu _{k}}$ is the generalized totally
anti-symmetric Kronecker delta and $R_{\mu \nu }^{\phantom{\mu\nu}{\rho%
\sigma}}$ is the Riemann tensor. It is worthwhile to mention that in $d$
dimensions, all terms for which $k>[d/2]$ are identically equal to zero, and
the term $k=d/2$ is a topological term. So, only terms for which $k<d/2$ are
contributing to the field equations.

In this paper we want to restrict ourself to the first four terms of
Lovelock gravity. The first term is the cosmological term which we ignore it
in the investigation of the properties and thermodynamics of the solutions.
The second term is the Einstein term, and the third and fourth terms are the
second order Lovelock (Gauss-Bonnet) and third order Lovelock terms
respectively. From a geometric point of view, the combination of these terms
in seven-dimensional spacetimes, is the most general Lagrangian producing
second order field equations, as in the four-dimensional gravity which the
Einstein-Hilbert action is the most general Lagrangian producing second
order field equations. Here, we will obtain asymptotically flat solutions of
third order Lovelock gravity and investigate their thermodynamics.

Indeed, it is interesting to explore black holes in generalized gravity
theories in order to discover which properties are peculiar to Einstein's
gravity, and which are robust features of all generally covariant theories
of gravity. Due to the nonlinearity of the field equations, it is very
difficult to find out nontrivial exact analytical solutions of Einstein's
equation with the higher curvature terms. In most cases, one has to adopt
some approximation methods or find solutions numerically. However, exact
static spherically symmetric black hole solutions of the Gauss-Bonnet
gravity have been found in Ref. \cite{Des,Whe}, and of the
Einstein-Maxwell-Gauss-Bonnet and Einstein-Born-Infeld-Gauss-Bonnet models
in Refs. \cite{Wil1,Wil2}. Black hole solutions with nontrivial topology in
this theory have been also studied in Refs. \cite{Cai,Ish}. The
thermodynamics of the uncharged static spherically black hole solutions has
been considered in \cite{MS} and of charged solutions in \cite{Wil2,Odin}.
All of these known solutions in Gauss-Bonnet gravity are static. Recently
one of us has introduced two new classes of rotating solutions of second
order Lovelock gravity and investigate their thermodynamics \cite{Deh1}.

Also the static spherically symmetric solutions of the
dimensionality continued gravity have been explored in Ref.
\cite{Zan}, while black hole solutions with nontrivial topology in
this theory have been studied in Ref. \cite{Aros}. The
thermodynamics of these solutions have been investigated in Refs.
\cite{Mun,Chr,Clu}. Up to our knowledge, no asymptotically flat
solution for Lovelock gravity higher than second order
(Gauss-Bonnet) has been obtained till now with two or more
fundamental constants. Indeed, the asymptotically flat solution of
Ref. \cite{Chr} has only one fundamental constant which is the
gravitational constant $G$. In this paper we want to find new
static solutions of third order Lovelock gravity which are
asymptotically flat and contain two and three fundamental
constants and investigate their thermodynamics. As we will show
later, these asymptotically flat solutions have some properties
which do not occur in Einstein or Gauss-Bonnet gravity.

The outline of our paper is as follows. We give a brief review of the field
equations of third order Lovelock gravity in Sec. \ref{Fiel}. In Sec. \ref
{Sol}, we present the static solutions of third order Lovelock gravity in
the presence of electromagnetic field with special values of $\alpha_2$ and $%
\alpha_3$, and investigate their properties. In Sec. \ref{Therm} we obtain
mass, entropy, temperature, charge, and electric potential of the $d$%
-dimensional black hole solutions and show that these quantities
satisfy the first law of thermodynamics. We also perform a local
stability analysis of the black holes in the canonical and grand
canonical ensembles. In Sec. \ref {Gen}, we introduce the general
asymptotically flat solutions with three fundamental constants and
investigate their thermodynamics. We finish our paper with some
concluding remarks.

\section{Field equations\label{Fiel}}

The main fundamental assumptions in standard general relativity are the
requirements of general covariance and that the field equations for the
metric be second order. Based on the same principles, the Lovelock
Lagrangian is the most general Lagrangian in classical gravity which
produces second order field equations for the metric. The action of third
order Lovelock gravity in the presence of electromagnetic field may be
written as
\begin{equation}
I=\int d^{d}x\sqrt{-g}\left( -2\Lambda +\mathcal{L}_{1}+\alpha _{2}\mathcal{L%
}_{2}+\alpha _{3}\mathcal{L}_{3}-F_{\mu \nu }F^{\mu \nu }\right)
\label{Act1}
\end{equation}
where $\Lambda $ is the cosmological constant, $\alpha _{2}$ and $\alpha
_{3} $ are Gauss-Bonnet and third order lovelock coefficients, $\mathcal{L}%
_{1}=R$ is just the Einstein-Hilbert Lagrangian, $\mathcal{L}_{2}=R_{\mu \nu
\gamma \delta }R^{\mu \nu \gamma \delta }-4R_{\mu \nu }R^{\mu \nu }+R^{2}$
is the Gauss-Bonnet Lagrangian, and

\begin{eqnarray}
\mathcal{L}_{3} &=& 2R^{\mu \nu \sigma \kappa }R_{\sigma \kappa \rho \tau
}R_{\phantom{\rho \tau }{\mu \nu }}^{\rho \tau }+8R_{\phantom{\mu
\nu}{\sigma \rho}}^{\mu \nu }R_{\phantom {\sigma \kappa} {\nu \tau}}^{\sigma
\kappa }R_{\phantom{\rho \tau}{ \mu \kappa}}^{\rho \tau }+24R^{\mu \nu
\sigma \kappa }R_{\sigma \kappa \nu \rho }R_{\phantom{\rho}{\mu}}^{\rho }
\nonumber \\
&&+3RR^{\mu \nu \sigma \kappa }R_{\sigma \kappa \mu \nu }+24R^{\mu \nu
\sigma \kappa }R_{\sigma \mu }R_{\kappa \nu }+16R^{\mu \nu }R_{\nu \sigma
}R_{\phantom{\sigma}{\mu}}^{\sigma }-12RR^{\mu \nu }R_{\mu \nu }+R^{3}
\label{L3}
\end{eqnarray}
is the third order Lovelock Lagrangian. In Eq. (\ref{Act1}) $F_{\mu \nu
}=\partial _{\mu }A_{\nu }-\partial _{\nu }A_{\mu }$ is electromagnetic
tensor field and $A_{\mu }$ is the vector potential.

Since in Lovelock gravity, only terms for which $k<d/2$ are contributing to
the field equations and we want to consider the third order lovelock
gravity, therefore we consider the $d$-dimensional spacetimes with $d\geq 7$%
. Varying the action with respect to the metric tensor $g_{\mu \nu }$ and
electromagnetic tensor field $F_{\mu \nu }$ the equations of gravitation and
electromagnetic fields are obtained as:
\begin{eqnarray}
&&G_{\mu \nu }^{(1)}+\Lambda g_{\mu \nu }+\alpha _{2}G_{\mu \nu
}^{(2)}+\alpha _{3}G_{\mu \nu }^{(3)}=T_{\mu \nu }  \label{Geq} \\
&&\nabla _{\nu }F^{\mu \nu }=0  \label{EMeq}
\end{eqnarray}
where $T_{\mu \nu }=2F_{\phantom{\lambda}{\mu}}^{\rho }F_{\rho \nu }-\frac{1%
}{2}F_{\rho \sigma }F^{\rho \sigma }g_{\mu \nu }$ is the energy-momentum
tensor of electromagnetic field, $G_{\mu \nu }^{(1)}$ is just the Einstein
tensor, and $G_{\mu \nu }^{(2)}$ and $G_{\mu \nu }^{(3)}$ are given as \cite
{Hoi}:
\begin{eqnarray*}
G_{\mu \nu }^{(2)} &=&2(-R_{\mu \sigma \kappa \tau }R_{\phantom{\kappa \tau
\sigma}{\nu}}^{\kappa \tau \sigma }-2R_{\mu \rho \nu \sigma }R^{\rho \sigma
}-2R_{\mu \sigma }R_{\phantom{\sigma}\nu }^{\sigma }+RR_{\mu \nu })-\frac{1}{%
2}\mathcal{L}_{2}g_{\mu \nu } \\
G_{\mu \nu }^{(3)} &=&-3(4R^{\tau \rho \sigma \kappa }R_{\sigma \kappa
\lambda \rho }R_{\phantom{\lambda }{\nu \tau \mu}}^{\lambda }-8R_{%
\phantom{\tau \rho}{\lambda \sigma}}^{\tau \rho }R_{\phantom{\sigma
\kappa}{\tau \mu}}^{\sigma \kappa }R_{\phantom{\lambda }{\nu \rho \kappa}%
}^{\lambda }+2R_{\nu }^{\phantom{\nu}{\tau \sigma \kappa}}R_{\sigma \kappa
\lambda \rho }R_{\phantom{\lambda \rho}{\tau \mu}}^{\lambda \rho } \\
&&-R^{\tau \rho \sigma \kappa }R_{\sigma \kappa \tau \rho }R_{\nu \mu }+8R_{%
\phantom{\tau}{\nu \sigma \rho}}^{\tau }R_{\phantom{\sigma \kappa}{\tau \mu}%
}^{\sigma \kappa }R_{\phantom{\rho}\kappa }^{\rho }+8R_{\phantom
{\sigma}{\nu \tau \kappa}}^{\sigma }R_{\phantom {\tau \rho}{\sigma \mu}%
}^{\tau \rho }R_{\phantom{\kappa}{\rho}}^{\kappa } \\
&&+4R_{\nu }^{\phantom{\nu}{\tau \sigma \kappa}}R_{\sigma \kappa \mu \rho
}R_{\phantom{\rho}{\tau}}^{\rho }-4R_{\nu }^{\phantom{\nu}{\tau \sigma
\kappa }}R_{\sigma \kappa \tau \rho }R_{\phantom{\rho}{\mu}}^{\rho
}+4R^{\tau \rho \sigma \kappa }R_{\sigma \kappa \tau \mu }R_{\nu \rho
}+2RR_{\nu }^{\phantom{\nu}{\kappa \tau \rho}}R_{\tau \rho \kappa \mu } \\
&&+8R_{\phantom{\tau}{\nu \mu \rho }}^{\tau }R_{\phantom{\rho}{\sigma}%
}^{\rho }R_{\phantom{\sigma}{\tau}}^{\sigma }-8R_{\phantom{\sigma}{\nu \tau
\rho }}^{\sigma }R_{\phantom{\tau}{\sigma}}^{\tau }R_{\mu }^{\rho }-8R_{%
\phantom{\tau }{\sigma \mu}}^{\tau \rho }R_{\phantom{\sigma}{\tau }}^{\sigma
}R_{\nu \rho }-4RR_{\phantom{\tau}{\nu \mu \rho }}^{\tau }R_{\phantom{\rho}%
\tau }^{\rho } \\
&&+4R^{\tau \rho }R_{\rho \tau }R_{\nu \mu }-8R_{\phantom{\tau}{\nu}}^{\tau
}R_{\tau \rho }R_{\phantom{\rho}{\mu}}^{\rho }+4RR_{\nu \rho }R_{%
\phantom{\rho}{\mu }}^{\rho }-R^{2}R_{\nu \mu })-\frac{1}{2}\mathcal{L}%
_{3}g_{\mu \nu }
\end{eqnarray*}

\section{Static Solutions \label{Sol}}

The metric of $d$-dimensional static spherically symmetric spacetime and the
vector potential may be written as:
\begin{eqnarray}
ds^{2} &=&-f(r)dt^{2}+\frac{dr^{2}}{f(r)}+r^{2}d\Omega ^{2}  \nonumber \\
A_{\mu } &=&h(r)\delta _{\mu }^{t}  \label{met}
\end{eqnarray}
where $d\Omega ^{2}$ is the metric of a $(d-2)$-sphere. The functions $f(r)$
and $h(r)$ may be obtained by solving the field equations (\ref{Geq}) and (%
\ref{EMeq}).

\subsection{Seven-dimensional Solutions}

As stated before, the gravitational field equation of third order Lovelock
gravity in seven dimensions is the most general second order differential
equation which presents the solutions of gravity. Indeed, the solution of
third order Lovelock gravity in seven dimensions is the most general
solution of gravity, based on the principle of standard general relativity.
Therefore, first we obtain the seven-dimensional solutions of third order
Lovelock gravity in the presence of electromagnetic field and investigate
their properties. Using Eq. (\ref{EMeq}) one can show that $h(r)=q/(4r^{4})$%
, where $q$ is an arbitrary real constant which is related to the charge of
the solution. To find the function $f(r)$, one may use any components of Eq.
(\ref{Geq}). The simplest equation is the $rr$-component of these equations
which can be written as
\begin{equation}
\left( -3\alpha _{3}(f-1)^{2}+\alpha _{2}(f-1)r^{2}-\frac{r^{4}}{24}\right)
r^{5}f^{\prime }+\alpha _{2}r^{6}(f-1)^{2}-\frac{r^{8}}{6}(f-1)-\frac{%
\Lambda }{60}r^{10}=\frac{q^{2}}{60}  \label{Eqf7}
\end{equation}
where prime denotes the derivative with respect to $r$. We consider the
solutions of Eq. (\ref{Eqf7}) for any arbitrary values of $\alpha _{2}$ and $%
\alpha _{3}$ in Sec. \ref{Gen}. Here, we study the special case of $\alpha
_{3}=2{\alpha _{2}}^{2}=\alpha ^{2}/72$. Even for this special case, there
exist three fundamental constants, $\Lambda $, $G=1$ and $\alpha $ in the
solution, while the solution of ref. \cite{Chr} has two fundamental
constants. As we will see below this solution has some properties which will
not happen in the Gauss-Bonnet gravity with three fundamental constants. The
real solution of Eq. (\ref{Eqf7}) with the above values of $\alpha _{2}$ and
$\alpha _{3}$ is
\begin{equation}
f(r)=1+{\frac{{r}^{2}}{\alpha }}\left\{ 1-\left( {1+\frac{\Lambda \alpha }{5}%
+\frac{3\alpha m}{5r^{6}}-\frac{3\alpha {q}^{2}}{10r^{10}}}\right)
^{1/3}\right\}  \label{F7sp}
\end{equation}
In the above equations $m$ is an integration constant which is related to
the mass of the solution. Unlike the solutions in Gauss-Bonnet gravity which
have two branches, here the solution (\ref{F7sp}) has only one branch.
Indeed, Eq. (\ref{Eqf7}) with the above $\alpha$'s has the real solution (%
\ref{F7sp}) and two complex solutions which are the complex
conjugate of each other. This feature is the same as the
asymptotically anti-de Sitter (AdS) solution of Ref. \cite {Chr}
which has a unique anti de Sitter vacuum.

In order to study the general structure of this solution, we first
look for the asymptotic behavior of the solutions. It is easy to
find out that these solutions are asymptotically flat for
$\Lambda=0$, AdS for $\Lambda<0$ and de Sitter (dS) for
$\Lambda>0$. In this paper we are interested in the case of
asymptotically flat solutions, and therefore we put $\Lambda=0$.
One can show that the Kretschmann scalar $R_{\mu \nu \lambda
\kappa }R^{\mu \nu \lambda \kappa }$ diverges at $r=0$, and
therefore there is a curvature singularity located at $r=0$. Now
we look for the existence of horizons, and therefore we look for
possible black hole solutions. The horizons, if any exist, are
given by the zeros of the function $f(r)=g^{rr}$.

First, we consider the uncharged solutions, where the horizon(s)
is located at
\begin{eqnarray}
r_{+} &=&\left\{ -\frac{\alpha }{2}+\frac{\sqrt{180m-75\alpha ^{2}}}{30}%
\right\} ^{1/2}  \label{horou} \\
r_{-} &=&\left\{ -\frac{\alpha }{2}-\frac{\sqrt{180m-75\alpha ^{2}}}{30}%
\right\} ^{1/2}  \label{horin}
\end{eqnarray}
As one can see from Eqs. (\ref{horou}), and (\ref{horin}), for
$\alpha >0$ there exists only one horizon provided the mass
parameter, $m$, is greater than $m_{\mathrm{crit}}=5/3\alpha
^{2}$. That is, there exist a minimum value for the mass (or
radius of horizon) in order to have uncharged asymptotically flat
black hole solution. More interesting is the uncharged solutions
of third order Lovelock gravity with $\alpha <0$. In this case
there exist an extreme value for the mass parameter, $m_{\mathrm{ext}%
}=5/12\alpha ^{2}$. The uncharged solution presents a black hole with inner
and outer horizon provided $m>m_{\mathrm{ext}}$, an extreme black hole for $%
m=m_{\mathrm{ext}}$ and a naked singularity otherwise. This property happens
only for third order Lovelock gravity and does not occur in Einstein or
Gauss-Bonnet gravity. Thus, it is natural to suppose that new solutions in
higher order Lovelock gravity might provide us with a new window on some
corners of low energy limit of string theory.

Second, we consider the charged solutions. It is easy to show that the
solution presents a black hole solution with two inner and outer horizon,
provided the charge parameter $q$ is less than, $q_{\mathrm{ext}}$, an
extreme black hole for $q=q_{\mathrm{ext}}$ and a naked singularity
otherwise, where $q_{\mathrm{ext}}$ is
\begin{equation}
q_{\mathrm{ext}}=\frac{1}{240}\left\{5760 m^2+13200\alpha^2 m-7625
\alpha^4+5\alpha \sqrt{15(96m-25\alpha^2)^3} \right\}  \label{qext7}
\end{equation}

\subsection{$d$-dimensional Solutions}

The $d$-dimensional static solutions in third order Lovelock gravity may be
obtained by solving Eqs. (\ref{Geq}) and (\ref{EMeq}) for the metric given
in Eq. (\ref{met}). Using Eq. (\ref{EMeq}), one can show that $%
h(r)=q/[(d-3)r^{d-3}]$, where $q$ is an arbitrary real constant which is
related to the charge of the solution, and the function $f(r)$ is the
solution of the following equation:
\begin{eqnarray}
&&\left[ 180(_{\phantom{d}{5}}^{d-2})\alpha _{3}r(f-1)^{2}-6(_{\phantom{d}{3}%
}^{d-2})\alpha _{2}(f-1)r^{3}+\frac{d-2}{2}r^{5}\right] f^{\prime
}+\Lambda
r^{6}  \nonumber \\
&&+360(_{\phantom{d}{6}}^{d-2})\alpha _{3}(f-1)^{3}-12(_{\phantom{d}{4}%
}^{d-2})\alpha _{2}r^{2}(f-1)^{2}+(_{\phantom{d}{2}%
}^{d-2})r^{4}(f-1)=-q^{2}r^{10-2d}  \label{Eqf}
\end{eqnarray}
where prime denotes the derivative with respect to $r$. The solution of Eq. (%
\ref{Eqf}) for arbitrary values of $\alpha _{2}$ and $\alpha _{3}$ will be
introduced in Sec. \ref{Gen}. Here we consider the solutions of Eq. (\ref
{Eqf}) for $\alpha _{i}$'s given as
\begin{equation}
\alpha _{3}=\frac{\alpha ^{2}}{72(_{\phantom{d}{4}}^{d-3})},\hspace{1cm}%
\alpha _{2}=\frac{\alpha }{(d-3)(d-4)}  \label{con}
\end{equation}
Again, Eq. (\ref{Eqf}) with condition (\ref{con}) has one real and
two complex solutions which are the complex conjugate of each
other. The real solution of Eq. (\ref{Eqf}) with condition
(\ref{con}) is
\begin{equation}
f(r)=1+{\frac{{r}^{2}}{\alpha }}\left\{ 1-\left( {1+\frac{6\Lambda \alpha }{%
(d-1)(d-2)}+\frac{3\alpha m}{(d-2)r^{d-1}}-\frac{6\alpha {q}^{2}}{\left(
d-2\right) (d-3)r^{2d-4}}}\right) ^{1/3}\right\}  \label{fsp}
\end{equation}
where $m$ is the mass parameter.

These solutions are asymptotically flat for $\Lambda=0$ and AdS or
dS for negative or positive values of $\Lambda$ respectively. Also
one can show that the Kretschmann scalar diverges at $r=0$ and
therefore there is a curvature singularity located at $r=0$.

In order to investigate the existence of black hole solutions for $\Lambda
=0 $ we, first, consider the uncharged solutions. As in the case of
seven-dimensional solutions, one can show that for $\alpha >0$, there exists
a critical value of horizon radius, $r_{+\mathrm{crit}}$ (or a critical
value of mass, $m_{\mathrm{crit}}$) such that the solution presents a black
hole for $r_{+}>r_{+\mathrm{crit}}$, and a naked singularity otherwise,
where $r_{+\mathrm{crit}}$ is the larger real root of the following
equation:
\begin{equation}
3(d-3)r_{+}^{4}+3(d-5)\alpha r_{+}^{2}+(d-7)\alpha ^{2}=0  \label{mcrit}
\end{equation}
The critical mass, $m_{\mathrm{crit}}$ now may be obtained easily as
\begin{eqnarray}
m_{\mathrm{crit}}&=&\frac{(d-2)\left[ d+3+\sqrt{-3(d-1)(d-9)}\right] }{3}
\nonumber \\
&& \times \left\{ \frac{3d-15-\sqrt{-3(d-1)(d-9)}}{6}\right\}
^{(d-7)/2}\left( \frac{\alpha }{d-3}\right) ^{(d-3)/2}  \label{Ext0}
\end{eqnarray}
Note that the critical mass is real only for $d\leq 9$, and therefore there
exist no uncharged black hole in dimension greater than nine for $\alpha
_{2} $ and $\alpha _{3}$ given in Eqs. (\ref{con}). Unlike to this special
case, we will see in Sec. \ref{Gen} that black hole solutions exist in any
dimension for the arbitrary values of $\alpha _{2}$ and $\alpha _{3}$.

For $\alpha <0$, the $d$-dimensional solutions of third order Lovelock
gravity present a new property that does not occur in lower order of
Lovelock gravity. Indeed the uncharged solution for negative values of $%
\alpha $ presents a black hole with two inner and outer horizon, provided $%
r_{+}>r_{+\mathrm{ext}}$, an extreme black hole for $r_{+}=r_{+\mathrm{ext}}$%
, and a naked singularity otherwise, where\thinspace $r_{+\mathrm{ext}}$ is
the smaller solution of Eq. (\ref{Ext0}). In this case the extremal value of
mass is
\begin{eqnarray}
m_{\mathrm{ext}}&=&\frac{(d-2)\left[ d+3-\sqrt{-3(d-1)(d-9)}\right] }{3}
\nonumber \\
&& \left\{ \frac{3d-15+\sqrt{-3(d-1)(d-9)}}{6}\right\} ^{(d-7)/2}\left(
\frac{\left| \alpha \right| }{d-3}\right) ^{(d-3)/2}  \label{mext}
\end{eqnarray}
Second, we consider the charged solutions. It is easy to show that the
solution presents a black hole solution with two inner and outer horizon,
provided $q$ is less than $q_{\mathrm{ext}}$, an extreme black hole for $%
q=q_{\mathrm{ext}}$, and a naked singularity otherwise , where $q_{\mathrm{%
ext}}$ is the solution of
\begin{equation}
3(d-5)\alpha r_{+}^{2d-8}+3(d-3)r_{+}^{2d-6}+(d-7)\alpha
^{2}r_{+}^{2d-10}-6(d-2)^{-1}q^{2}=0  \label{Ext}
\end{equation}

\section{Thermodynamics of black holes \label{Therm}}

One can obtain the temperature of the event horizon by analytic continuation
of the metric. The analytical continuation of the Lorentzian metric by $%
t\rightarrow i\tau $ yields the Euclidean section, whose regularity at $%
r=r_{+}$ requires that we should identify $\tau \sim \tau +\beta _{+}$,
where $\beta _{+}$ is the inverse Hawking temperature of the horizon given
as
\begin{equation}
\beta _{+}^{-1}=T_{+}=\frac{f^{\prime }(r)}{4\pi }=\frac{%
(d-2)r_{+}^{2d-10}[3(d-3)r_{+}^{4}+3(d-5)\alpha r_{+}^{2}+(d-7)\alpha
^{2}]-2q^{2}}{12\pi (r_{+}^{2}+\alpha )^{2}r_{+}^{2d-9}}  \label{Temp}
\end{equation}

Usually entropy of the black holes satisfies the so-called area law of
entropy which states that the black hole entropy equals to one-quarter of
horizon area \cite{Beck}. One of the surprising and impressive feature of
this area law of entropy is its universality. It applies to all kind of
black holes and black strings of Einstein gravity \cite{Haw}. However, in
higher derivative gravity the area law of entropy is not satisfied in
general \cite{fails}. It is known that the entropy in Lovelock gravity is
\cite{Myer}
\begin{equation}
S=\frac{1}{4} \sum_{k=1}^{[(d-1)/2]}k\alpha _{k}\int d^{d-2}x \sqrt{\tilde{g}%
} \tilde{\mathcal{L}}_{k-1}  \label{Enta}
\end{equation}
where the integration is done on the $(d-2)$-dimensional spacelike
hypersurface of Killing horizon, $\tilde{g}_{\mu \nu }$ is the induced
metric on it, $\tilde{g}$ is the determinant of $\tilde{g}_{\mu \nu }$ and $%
\tilde{\mathcal{L}}_{k}$ is the $k$th order Lovelock Lagrangian of $\tilde{g}%
_{\mu \nu } $. Thus, the entropy in third order Lovelock gravity is
\begin{equation}
S=\frac{1}{4} \int d^{d-2}x \sqrt{\tilde{g}}\left( 1+2\alpha _{2}\tilde{R}%
+3\alpha _{3}(\tilde{R}_{\mu \nu \sigma \kappa }\tilde{R}^{\mu \nu \sigma
\kappa }-\tilde{R}_{\mu \nu }\tilde{R}^{\mu \nu }+\tilde{R}^{2})\right)
\label{Entb}
\end{equation}
where $\tilde{R}_{\mu \nu \rho \sigma }$ and $\tilde{R}_{\mu \nu }$ are
Riemann and Ricci tensors and $\tilde{R}$ is the Ricci scalar for the
induced metric $\tilde{g}_{ab}$ on the $(d-2)$-dimensional horizon. It is a
matter of calculation to show that the entropy of black holes is
\begin{equation}
S=\frac{1}{4}r_{+}^{d-2}\Sigma _{d-2}\left( 1+\frac{4\alpha _{2}(_{%
\phantom{d}{2}}^{d-2})}{r_{+}^{2}}+\frac{72(_{\phantom{d}{4}}^{d-2})\alpha
_{3}}{r_{+}^{4}}\right)  \label{Ent1}
\end{equation}
where $\Sigma _{d-2}$ is the area of a unit radius $(d-2)$-dimensional
sphere.

The charge of the black hole can be found by calculating the flux of the
electric field at infinity, yielding
\begin{equation}
Q =\frac{\Sigma _{d-2}}{4\pi }q  \label{Ch}
\end{equation}
The electric potential $\Phi $, measured at infinity with respect to the
horizon, is defined by \cite{Cal}
\begin{equation}
\Phi =A_{\mu }\chi ^{\mu }\left| _{r\rightarrow \infty }-A_{\mu }\chi ^{\mu
}\right| _{r=r_{+}},  \label{Pot1}
\end{equation}
where $\chi =\partial /\partial t$ is the null generator of the horizon. One
finds
\begin{equation}
\Phi =\frac{q}{(d-3)r_{+}^{d-3}}  \label{Pot2}
\end{equation}

The mass of black hole can be obtained by using the behavior of the metric
at large $r$. It is easy to show that the mass of black hole is
\begin{equation}
M=\frac{\Sigma _{d-2}}{16\pi }m=\frac{(d-2)(d-3)(3r_{+}^{4}+3\alpha
r_{+}^{2}+\alpha ^{2})r_{+}^{2d-10}+2q^{2}}{(d-3)r_{+}^{d-3}}  \label{Mass1}
\end{equation}

We now investigate the first law of thermodynamics. Using the expression for
the entropy, the charge and the mass given in Eqs. (\ref{Ent1}), (\ref{Ch})
and (\ref{Mass1}), one can compute $\partial M/\partial r$, $\partial
S/\partial r$ and $\partial Q/\partial r$. Then, by using the chain rule, it
is easy to show that the quantities
\begin{equation}
T=\left( \frac{\partial M}{\partial S}\right) _{Q},\ \ \ \ \Phi =\left(
\frac{\partial M}{\partial Q}\right) _{S}  \label{Dsmar}
\end{equation}
are exactly the same as the temperature and electric potential given in Eqs.
(\ref{Temp}) and (\ref{Pot2}) respectively. Thus, the thermodynamic
quantities calculated in Eqs. (\ref{Temp}) and (\ref{Pot2}) satisfy the
first law of thermodynamics,
\begin{equation}
dM=TdS+\Phi dQ  \label{1stlaw}
\end{equation}

\subsection{Stability in the canonical and the grand-canonical ensemble}

The stability of a thermodynamic system with respect to the small variations
of the thermodynamic coordinates, is usually performed by analyzing the
behavior of the entropy $S(M,Q)$ around the equilibrium. The local stability
in any ensemble requires that $S(M,Q)$ be a convex function of their
extensive variables or its Legendre transformation must be a concave
function of their intensive variables. Thus, the local stability can in
principle be carried out by finding the determinant of the Hessian matrix of
$S$ with respect to its extensive variables, $\mathbf{H}_{X_{i}X_{j}}^{S}=[%
\partial ^{2}S/\partial X_{i}\partial X_{j}]$, where $X_{i}$'s are the
thermodynamic variables of the system. Indeed, the system is locally stable
if the determinant of Hessian matrix satisfies $\mathbf{H}%
_{X_{i},X_{j}}^{S}\leq 0$ \cite{Cvet}. Also, one can perform the stability
analysis through the use of the determinant of Hessian matrix of the energy
with respect to its thermodynamic variables, and the stability requirement $%
\mathbf{H}_{X_{i},X_{j}}^{S}\leq 0$ may be rephrased as $\mathbf{H}%
_{Y_{i},Y_{j}}^{M}\geq 0$ \cite{Gub}.

The number of the thermodynamic variables depends on the ensemble which is
used. In the canonical ensemble, the charge is a fixed parameter, and
therefore the positivity of the heat capacity $C_{Q}=T(\partial S/\partial
T)_{Q}$ is sufficient to assure the local stability. The heat capacity for
the solutions given by Eq. (\ref{fsp}) shows that there exist an upper limit
for the radius of horizon, $r_{+\mathrm{st}}$ which is the real root of $St_{%
\mathrm{can}}=0$, where $St_{\mathrm{can}}$ is
\begin{eqnarray}
St_{\mathrm{can}}&=& ( d-2 )[( d-7 ) {\alpha}^{3}+2{r_{+\mathrm{st}}}^{2} (
d-10 ) {\alpha}^{2}-18{r_{+\mathrm{st}}}^{4}\alpha+3{r_{+\mathrm{st}}}^{6} (
d-3 )]  \nonumber \\
&&-6( 2d-5 ) {r_{+\mathrm{st}}}^{-2d+12}{q}^{2}-6(2d-9){r_{+\mathrm{st}}}%
^{-2d+10}\alpha {q}^{2}  \label{Stcan}
\end{eqnarray}
Therefore, the charged solutions are stable provided the horizon radius lies
in the range $r_{+\mathrm{ext}}<r_{+}<r_{+\mathrm{st}}$, where $r_{+\mathrm{%
ext}}$ is the solution of Eq. (\ref{Ext}), and $r_{+\mathrm{st}}$ is the
real solution of Eq. (\ref{Stcan}). Indeed, for $r_{+}<r_{+\mathrm{ext}}$,
we have no black hole while for $r_{+}>r_{+\mathrm{st}}$, the black hole is
not stable and therefore we have only an intermediate stable phase. Note
that for the case of uncharged black holes and $\alpha >0$, $r_+$ should be
greater than $r_{+crit}$ as stated before. Thus for positive $\alpha$, the
horizon radius of stable black holes lies in the range $r_{+\mathrm{crit}%
}<r_{+}<r_{+\mathrm{st}}$, where $r_{+\mathrm{crit}}$ is the larger root of
Eq. (\ref{Ext0}). Of course, for negative $\alpha $, which one can have a
black hole with inner and outer horizon, the black hole is stable provided $%
r_+$ lies in the range $r_{+\mathrm{ext}}<r_{+}<r_{+\mathrm{st}}$, where now
$r_{+\mathrm{ext}}$ is the smaller root of Eq. (\ref{Ext0}), and again we
have only an intermediate stable phase.

In the grand-canonical ensemble, the stability analysis can be carried out
by calculating the determinant of Hessian matrix of the energy with respect
to $S$ and $Q$. The zeros of the determinant of Hessian matrix are given by $%
St_{\mathrm{gc}}=0$, where $St_{\mathrm{gc}}$ is
\begin{eqnarray}
St_{\mathrm{gc}} &=&(d-2)[(d-7){\alpha }^{3}+2\,{r}_{+\mathrm{st}}^{2}(d-10){%
\alpha }^{2}-18\,{r}_{+\mathrm{st}}^{4}\alpha +3\,{r}_{+\mathrm{st}%
}^{6}(d-3)]  \nonumber \\
&&+18\,{r}_{+\mathrm{st}}^{-2\,d+10}{q}^{2}\alpha -6\,{r}_{+\mathrm{st}%
}^{-2\,d+12}{q}^{2}  \label{Stgc}
\end{eqnarray}
Three cases happen for the roots of $St_{\mathrm{gc}}=0$:\newline
\noindent 1. It has two real roots $r_{+\mathrm{st1}}<r_{+\mathrm{st2}}$. In
this case, the determinant of Hessian matrix is positive provided the radius
of horizon lies in the range between $r_{+\mathrm{larger}}<r_{+}<r_{+\mathrm{%
st2}}$, where $r_{+\mathrm{larger}}$ is the larger value between $r_{+%
\mathrm{ext}}$ and $r_{+\mathrm{st1}}$. Thus, there exist an intermediate
stable phase\newline
\noindent 2. It has one real root $r_{+\mathrm{st2}}$. In this case, there
exists an intermediate stable phase with radius of horizon between $r_{+%
\mathrm{ext}}$ and $r_{+\mathrm{st2}}$. \newline
\noindent 3. It has no real solution. In this case the black hole is
unstable for the whole range of $r_{+}$.

Numerical analysis shows that $r_{+\mathrm{st2}}<r_{+\mathrm{st}}$ (if $r_{+%
\mathrm{st2}}$ exists), and therefore the region of stability is
smaller for the grand-canonical ensemble. This is due to the fact
that the number of thermodynamic variables in the canonical
ensemble is less than that of the grand-canonical ensemble.

\section{General Solutions \label{Gen}}

\begin{figure}[tbp]
\epsfxsize=10cm \centerline{\epsffile{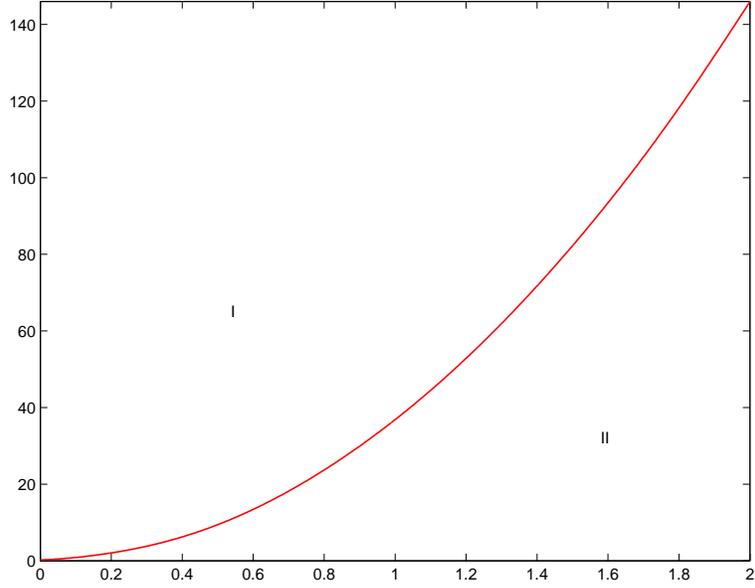}}
\caption{$m$ versus $\protect\alpha_2$ for $q=0$, $d=7$, and $\alpha%
_3=3\alpha_2/2$.} \label{Figure1}
\end{figure}

Finally we give the general solutions of third order Lovelock gravity in $d$
dimensions for any arbitrary values of $\alpha _{2}$ and $\alpha _{3}$. In
this case the solution of Eq. (\ref{Eqf}) is
\begin{equation}
f(r)=1+r^{2}\frac{(\frac{\alpha _{2}}{6\alpha _{3}})}{(_{\phantom{d}{2}%
}^{d-5})}+\frac{r^{\frac{-2(d-5)}{3}}}{\alpha _{3}}\left( \xi _{d}+\frac{%
\alpha _{3}}{24}\sqrt{\zeta _{d}}\right) ^{1/3}+\frac{r^{\frac{2(d+1)}{3}%
}\left( \frac{1}{(_{2}^{d-5})^{2}}\alpha _{2}^{2}-\frac{1}{2(_{\phantom{d}{4}%
}^{d-3})}\alpha _{3}\right) }{36\alpha _{3}\left( \xi _{d}+\frac{\alpha _{3}%
}{24}\sqrt{\zeta _{d}}\right) ^{1/3}}  \label{Fgen}
\end{equation}
where $\xi _{d}$ and $\zeta _{d}$ are
\[
\xi _{d}=\frac{\alpha _{2}^{3}r^{2(d-2)}}{216(_{\phantom{d}{2}}^{d-5})^{3}}-%
\frac{\alpha _{2}\alpha _{3}r^{2(d-2)}}{144(_{\phantom{d}{4}%
}^{d-3})(d-6)(d-5)}+\frac{m\alpha _{3}^{2}(d-2)r^{d-3}}{120(_{\phantom{d}{5}%
}^{d-2})}+\frac{\alpha _{3}^{2}q^{2}}{120(_{\phantom{d}{5}}^{d-2})(d-3)}
\]
\begin{eqnarray*}
\zeta _{d} &=&\frac{\alpha _{3}r^{4(d-2)}}{648(_{\phantom{d}{4}}^{d-3})^{3}}-%
\frac{\alpha _{2}^{2}r^{4(d-2)}}{36(d-6)^{2}(d-5)^{2}(_{\phantom{d}{4}%
}^{d-3})^{2}}+\frac{2^{\frac{1+(-1)^{d}}{2}}\alpha _{3}^{2}mq^{2}(d-2)r^{d-3}%
}{25(d-3)(_{\phantom{d}{5}}^{d-2})^{2}} \\
&&-\frac{\alpha _{2}\alpha _{3}m(d-2)r^{3d-7}}{15\times
2^{1-(-1)^{d}}(d-6)(d-5)(_{\phantom{d}{5}}^{d-2})(_{\phantom{d}{4}}^{d-3})}+%
\frac{\alpha _{3}^{2}m^{2}(d-2)^{2}r^{2(d-3)}}{25\times 4^{1-(-1)^{d}}(_{%
\phantom{d}{5}}^{d-2})^{2}} \\
&&-\frac{\alpha _{2}\alpha _{3}q^{2}r^{2(d-2)}}{3(_{\phantom{d}{4}}^{d-3})(_{%
\phantom{d}{4}}^{d-2})(d-3)(d-5)(d-6)^{2}}+\frac{8\times 2^{\frac{1+(-1)^{d}%
}{2}}(d-2)m\alpha _{2}^{3}r^{3d-7}}{45(d-5)^{3}(d-6)^{3}(_{\phantom{d}{5}%
}^{d-2})} \\
&&+\frac{q^{4}\alpha _{3}^{2}}{25(d-3)^{2}(_{\phantom{d}{5}}^{d-2})^{2}}+%
\frac{16q^{2}\alpha _{2}^{3}r^{2(d-2)}}{9(d-3)(d-2)(d-6)^{3}(d-5)^{3}(_{%
\phantom{d}{4}}^{d-3})}
\end{eqnarray*}
In the above equations $m$ is an integration constant which is
related to the mass of the solution. This solution is
asymptotically flat and has a curvature singularity at $r=0$. Note
that this asymptotically flat solution has three fundamental
constants $G=1$, $\alpha _{2}$ and $\alpha _{3}$. Numerical
analysis shows that the solutions (\ref{Fgen}) may present black
holes with two inner and outer horizons, extreme black holes,
black holes
with one event horizon, or naked singularities depending on the values of $%
\alpha _{2}$, $\alpha _{3}$, $m$ and $q$. To be more clear we give
the range of values of $m$, $\alpha _{2}$ and $\alpha _{3}$ for
the 7-dimensional uncharged solution in order to have black hole
solutions. For $\alpha _{3}=3\alpha _{2}/2$, any solution with $m$
and $\alpha _{2}$ in region I of Fig. \ref{Figure1} represents a
black hole solution with an event horizon.

\begin{figure}[tbp]
\epsfxsize=10cm \centerline{\epsffile{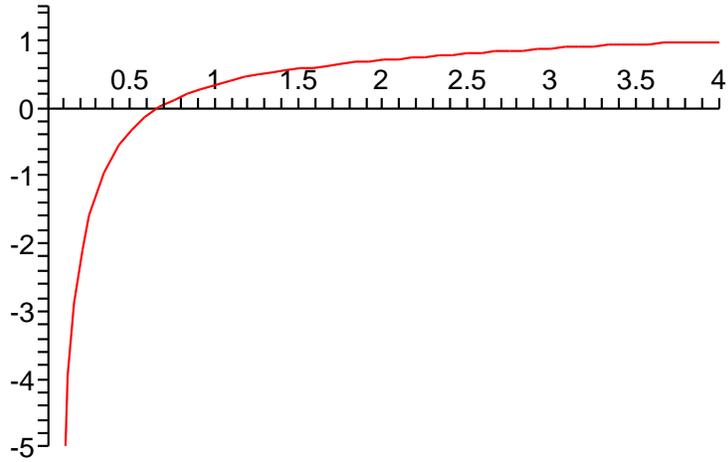}} \caption{$f(r)$
versus $r$ for $d=10$, $q=0$, $m=500$, $\alpha_2=1.0$, and
$\alpha_3=2.0$.} \label{Figure2}
\end{figure}

Unlike to the case of solutions with special values of $\alpha_2$ and $%
\alpha_3$ given in Sec. \ref{Sol}, the uncharged solution for arbitrary
values of fundamental constants $\alpha_2$ and $\alpha_3$ can have horizon
even for $d>9$. Figure \ref{Figure2} shows the function $f(r)$ for $d=10$ as
a function of $r$. It shows that there is an event horizon at $r=2$.

To investigate the first law of thermodynamics for the solutions with
arbitrary values of $\alpha _{2}$ and $\alpha _{3}$, one should perform a
numerical analysis. Using the expression for the entropy and the charge
given in Eqs. (\ref{Ent1}) and (\ref{Ch}), and the fact that $%
M=[\Sigma_{d-2}/(16\pi)]m$, where $m$ is the solution of $f(m,r_{+})=0$, one
can show numerically that the temperature and electric potential of Eq. (\ref
{Dsmar}) are exactly equal to the temperature and electric potential which
can be computed through the use of geometrical analysis. Thus, the first law
of thermodynamics is valid for the general solution given in Eq. (\ref{Fgen}%
). Also, one can perform a stability analysis in canonical and
grand-canonical ensembles numerically. Again, as in the case of
special solutions of third order Lovelock gravity, numerical
analysis shows that there exist only an intermediate stable phase
for black hole solutions which is smaller in grand-canonical
ensemble .

\section{CLOSING REMARKS}

In this paper, we added the third order Lovelock terms to the
Gauss-Bonnet-Maxwell action, and introduced a new class of static solutions
which are asymptotically flat, AdS or dS for $\Lambda=0$, $\Lambda<0$ or $%
\Lambda>0$ respectively. We were only interested in the
asymptotically flat solutions.

First, we consider the solutions for special values of $\alpha_2$ and $%
\alpha_3$ given in Eq. (\ref{con}). For uncharged solutions, we found that
for $\alpha>0$, the solutions present black holes provided the mass is
larger than a critical value, $m_{\mathrm{crit}}$, given in Eq. (\ref{mcrit}%
). For negative $\alpha$, they present black holes with two inner and outer
horizons for $m>m_{\mathrm{ext}}$, extreme black holes for $m=m_{\mathrm{ext}%
}$ or naked singularity for $m<m_{\mathrm{ext}}$, where
$m_{\mathrm{ext}}$ is given in Eq. (\ref{mext}). These kind of
uncharged solutions exist only in third order Lovelock gravity and
do not happen in Einstein or Gauss-Bonnet gravity. For the case of
charged solutions, we found that there exists an extremal value of
$r_+$, given by Eq. (\ref{Ext}), that determines whether the
solution presents a black holes with two inner and outer horizons,
an extreme black hole or a naked singularity. Accordingly we
obtained temperature, entropy, charge , electric potential and
mass of these black hole solutions. We also investigated the first
law of thermodynamics, and found that these thermodynamics
quantities satisfy the first law of thermodynamics. Also, we
performed a stability analysis in canonical ensemble by
considering the heat capacity of the solution and found that there
exist two unstable phases separated by an intermediate stable
phase. This analysis was also done through the use of the
determinant of the Hessian matrix of $M(S,Q)$ with respect to its
extensive variables and we got the same phase behavior with a
smaller region of stability in the grand-canonical ensemble or no
stable phase depending on the values of black hole's parameters.

Second, we introduced the general solution of third order Lovelock
gravity for arbitrary values of $\alpha_2$ and $\alpha_3$ and
investigate its properties. Numerical analysis showed that these
solutions may be interpreted as black hole solutions with two
inner and outer event horizons, extreme black holes or naked
singularity depending on the parameters of the solutions. We also
found that the conserved and thermodynamics quantities for these
general solutions satisfy first law of thermodynamics. Again, the
stability analysis showed that there exists only an intermediate
stable phase, for the black hole solutions which is smaller for
the grand-canonical ensemble.

As we mention, the asymptotically flat solutions obtained in
\cite{Chr} contain only one fundamental constant. Finding new
solutions in continued Lovelock gravity with more fundamental
constants remains to be carried out in future. Also, the
generalization of these solutions to the case of rotating
solutions will be given elsewhere.

\end{document}